\documentclass{ws-rv9x6}
\usepackage{subfigure}   
\usepackage{ws-rv-thm}   
\usepackage{ws-rv-van}   
\makeindex

\begin{document}

\chapter[Soft Interactions in Herwig]{Soft Interactions in Herwig}\label{ra_ch1}

\author[S.~Gieseke, P.~Kirchgae{\ss}er \& F.~Loshaj]{S.~Gieseke, P.~Kirchgae{\ss}er \& F.~Loshaj}

\address{Karlsruhe Institute of Technology, Institute for Theoretical Physics\\
Wolfgang-Gaede-Str.~1, 76131 Karlsruhe, Germany}

\begin{abstract}
  We introduce a new model for soft interactions in the Monte Carlo
  event generator Herwig.  We add a new model for the simulation of
  diffractive final states, based on the cluster hadronization model in
  Herwig.  The soft component of the Multiple partonic interaction model
  is replaced by a refined model for soft gluon production.  With these
  two components we are for the first time able to give a full
  simulation of minimum-bias events at hadron colliders.  We briefly
  describe the models and present a few results obtained with it. 
\end{abstract}


\body

\section{Introduction}
The underlying event plays an important r\^{o}le in the simulation of
particle collisions at hadron colliders.  The most important input for
the simulation of the underlying event is the usage of a multiple
partonic interaction (MPI) model that is capable to produce jets at
large transverse momenutum via the simulation of multiple semi-hard
partonic scatters \cite{Bahr:2008pv,Sjostrand:2007gs,Gleisberg:2008ta}.  

In Herwig, the MPI model has been implemented following the
\textsc{jimmy} model \cite{Butterworth:1996zw} and extending it with a
soft component \cite{Bahr:2009ek,Bahr:2008wk,Bahr:2008dy}.  Here, the
soft component adds additional gluons with low transverse momentum
according to a Gaussian in $p_\perp^2$ that is also found in
experimental data on the transverse momentum of very soft particles,
i.e.\ particles with transverse momenta lower that a few GeV.  These
soft gluons clearly have to be seen as a technical vehicle to simulate
soft physics rather than perturbative quanta of strong interactions.
They help us to connect the event on the parton level to the
hadronization stage of the event.  At this point, the colour structure
of the event, given by the connections of individual gluons to the
colour lines of partons from other events or the event remnants is most
important for the further evolution of the event.  Given the lack of a
first-principles approach to these soft interactions they were modeled
as a random assignment of either a colour connection as in a hard event,
i.e.\ with colour exchange, or as a peculiar event where the produced
gluons are disrupted from the colour structure of the remainder of the
event.  The weight towards one or the other structure was left as a
tuning parameter and fixed with a rather large probability to produce
colour disrupted events.

This modeling of soft events was intended as a smooth interpolation of
the hard events into the soft region with the main motivation to avoid
unphysical boundary effects in the simulation of MPI events, serving as
the underying event in hard events.  At the same time it also allowed
for a glimpse into the regime of minimum-bias (MB) events where no hard
scatter was present.  The model has turned out to be fairly stable
whenever a not too soft selection as a minimum transverse momentum of at
least 500\,MeV or the requirement of a handful of charged particles in
the event has been made.  At the same time, the model obviously failed
when it was applied to less restricted minimum-bias events.  This is
easily explained as a result of i) the rather poor modeling of soft
particle production and ii) the overall lack of a description of
diffractive events \cite{Gieseke:2016pbi}.  These two points have now
been addressed in our recent work \cite{Gieseke:2016fpz} that we report
on at this conference.

\begin{figure}
  \centering
  \includegraphics[width=.8\textwidth]{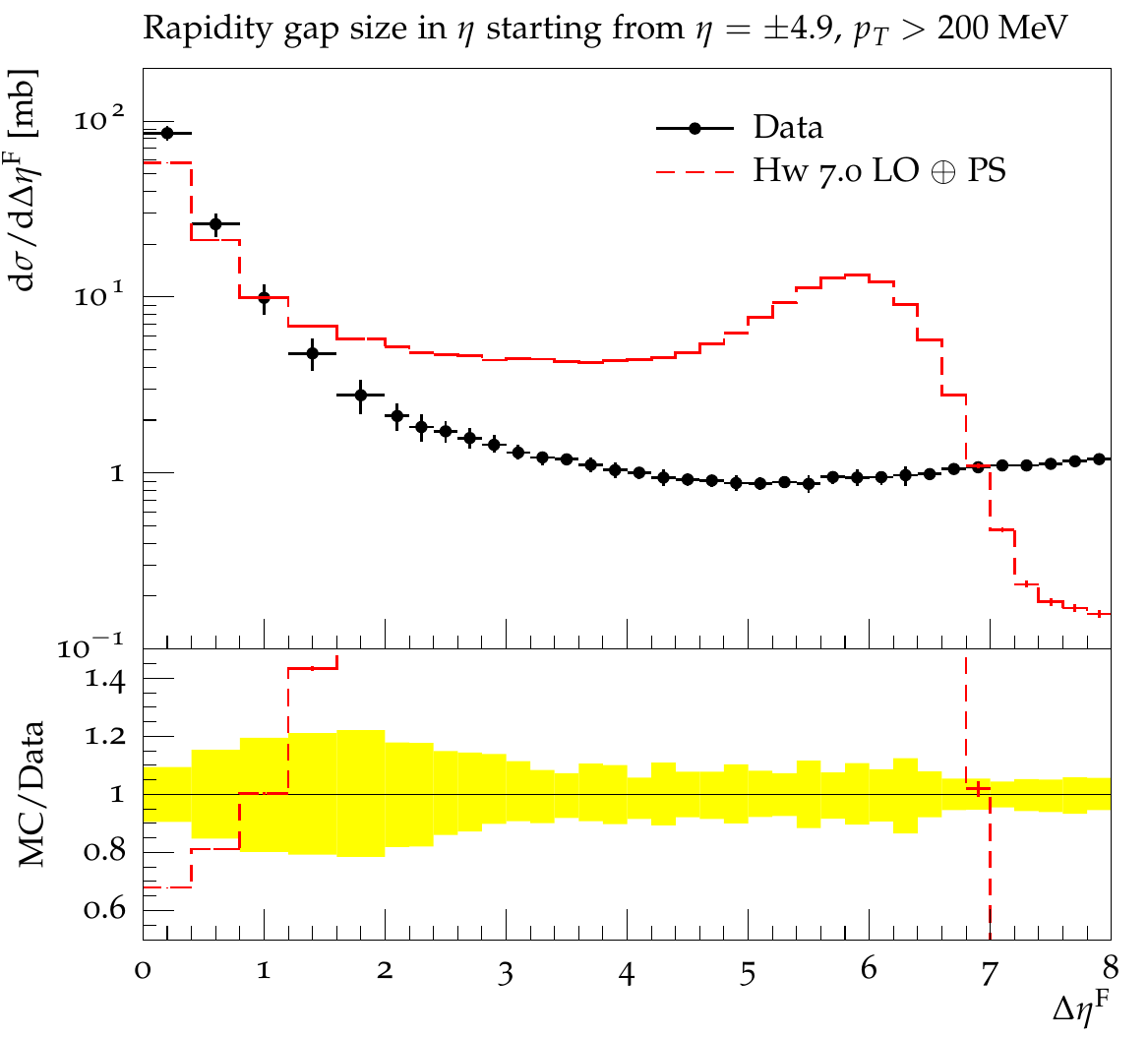}
  \caption{Distribution of events with large rapidity gap, as measured
    by ATLAS \cite{}.}
  \label{fig:bump}
\end{figure}

A particularly clear failure can be observed in Fig.~\ref{fig:bump}.
The observable $\Delta\eta_F$ describes the largest rapidity gap in the
detector from any given track towards the end of the detector
\cite{Aad:2012pw}.  This gap will be large for diffractive events where
the final state of one system disappears into the beam pipe while the
other system leaves tracks in the detector.  The observed bump in the
Herwig simulation is a result of e.g.\ events with disrupted colour
connections that result in large rapidity gaps.

\section{The new model for soft interactions}
This failure led us to further studies of the simulation of soft physics
with our event generator with the conclusion that we need to replace the
production of soft particles with a new, more sophisticated model and
that we need a model for diffractive final states to complement this
soft model.  

The latter model is, despite the hadronic nature of diffractive
interactions, formulated as a model of (non-perturbative) parton
production that can be used to produce low-mass diffractive final states
by means of the cluster hadronization model in Herwig.  We make use of a
Regge factorized ansatz for the diffractive cross section with an
exponentially falling $t$ dependence and a power-like $M$ dependence,
where $t, M$ are the commonly known momentum transfer and diffractive
mass of the system.  Once given these kinematics per event, we can
generate the production of a quark-diquark system with the given $t, M$
that will then hadronize into the desired diffractive final state.  For
very low masses we directly produce a $\Delta$ resonance.  

\begin{figure}
  \centering
  \includegraphics[width=\textwidth]{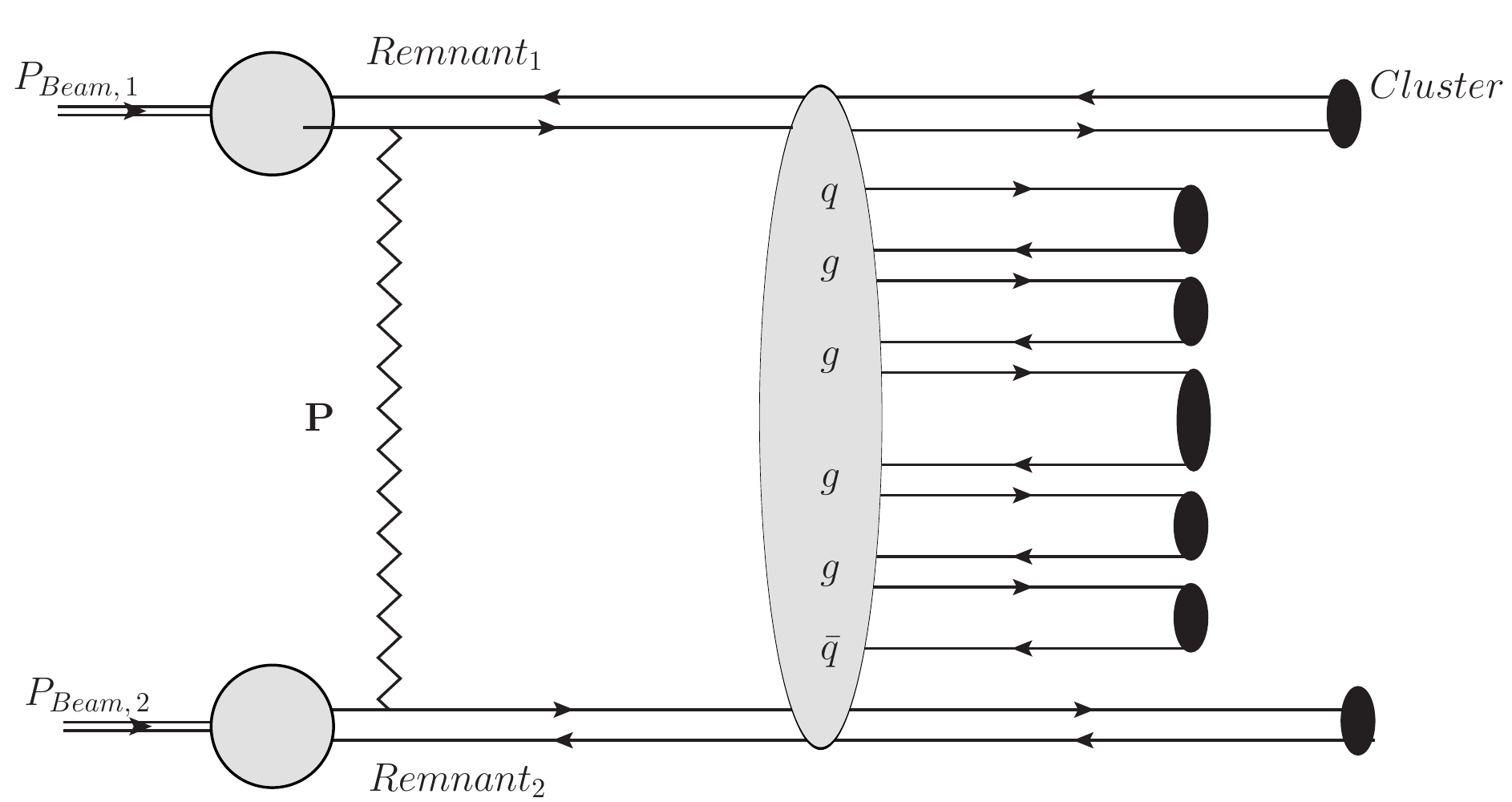}
  \caption{Colour connections of soft gluons in the ladder.}
  \label{fig:colourconn}
\end{figure}

\begin{figure}
  \centering
\includegraphics[width=.49\textwidth]{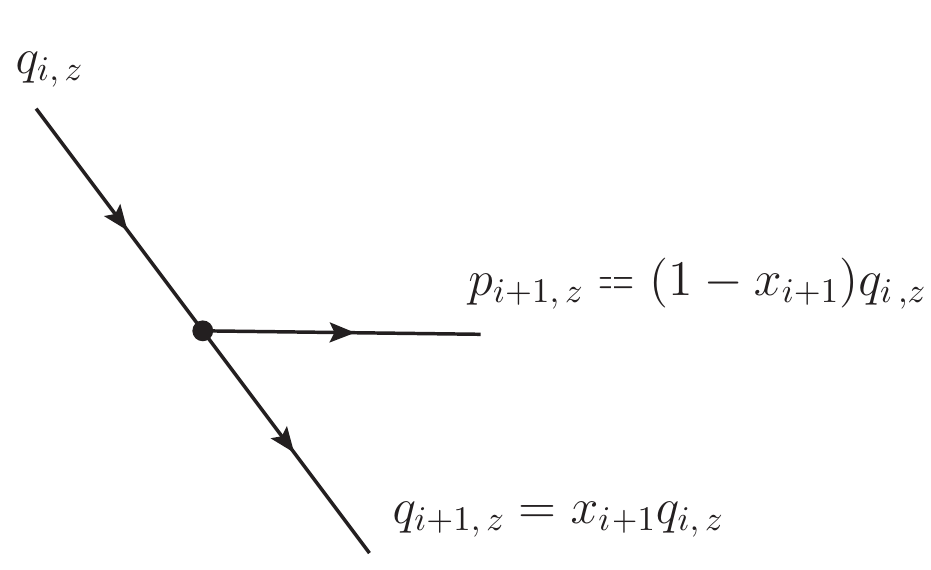}
  \hfill
  \includegraphics[width=.49\textwidth]{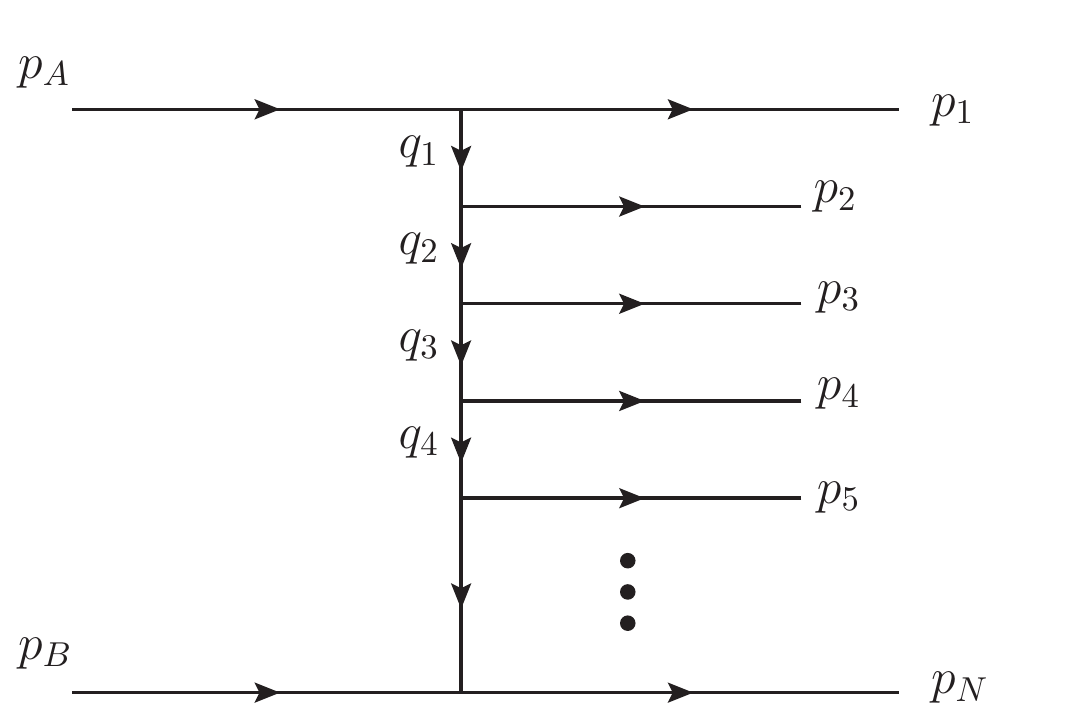}
  \caption{Kinematics of soft particle emissions in the ladder.}
  \label{fig:ladderkin}
\end{figure}

The model for soft particle production is much more detailed and is
based on the model for so-called multi-peripheral particle
production\cite{Baker:1976cv}.  Again, we translate the model into our
space of partonic degrees of freedom, where we produce soft gluons
approximately flat in rapidity, i.e.\ according to the multi-peripheral
kinematics.  In order to be able to hadronize this final state and to
avoid high-mass clusters in the system we have to add a quark pair to
the system and then produce the soft gluons ordered in rapidiy as
depicted in Fig.~\ref{fig:colourconn}.  One such ladder is the result of
one soft interaction according to our MPI model.  I.e\ where we
previously produced a single pair of gluons we now produce a whole
ladder of soft gluons cf.\ Fig.~\ref{fig:ladderkin}.  The number of
gluons depends on the available phase space in rapidity and is chosen to
keep the number of soft gluons approximately constant per rapidity
interval.  The transverse momenta are chosen according to a Gaussian, as
in the previous model for soft interactions.  A detailed description of the model is given in \cite{Gieseke:2016fpz}. 


\section{Results}
We have tuned the single parameter of the new model to minimum-bias data
at different centre-of-mass energies.  The embedding of the model takes
place entirely within the framework of the old MPI model, i.e.\ we still
have a mildly $\sqrt{s}$ dependent minimum transverse momentum and a
radius paremeter of the proton that both determine the number of soft
interactions, i.e. the number of ladders.  The number of particles
within the single ladder depends on the available rapidity span and
needs a single normalization factor that will as well mildly rise with
$\sqrt{s}$.  Furthermore we have to adjust the fraction of diffractive
events.  

\begin{figure}
  \centering
  \includegraphics[width=.49\textwidth]{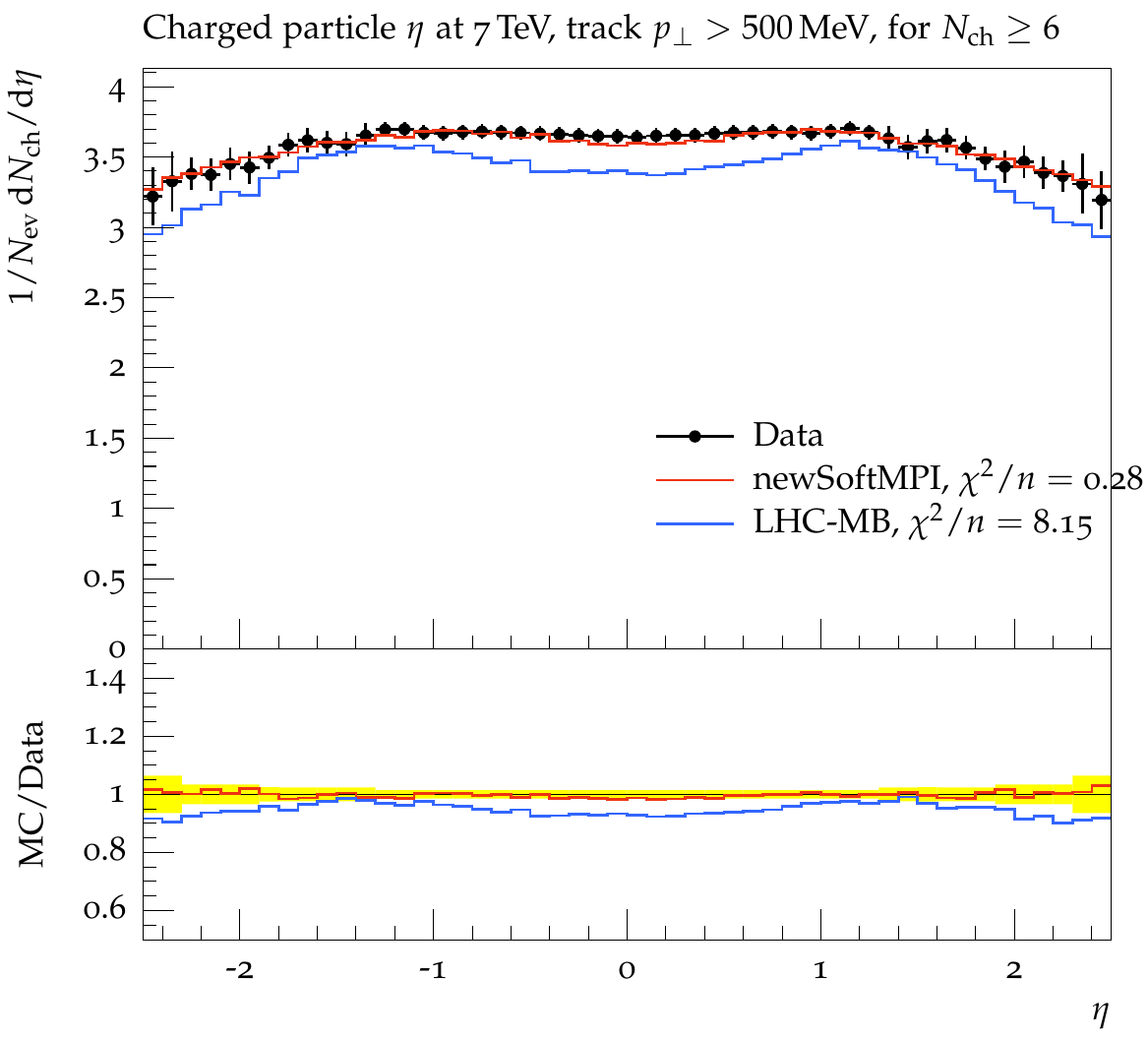}
  \hfill
  \includegraphics[width=.49\textwidth]{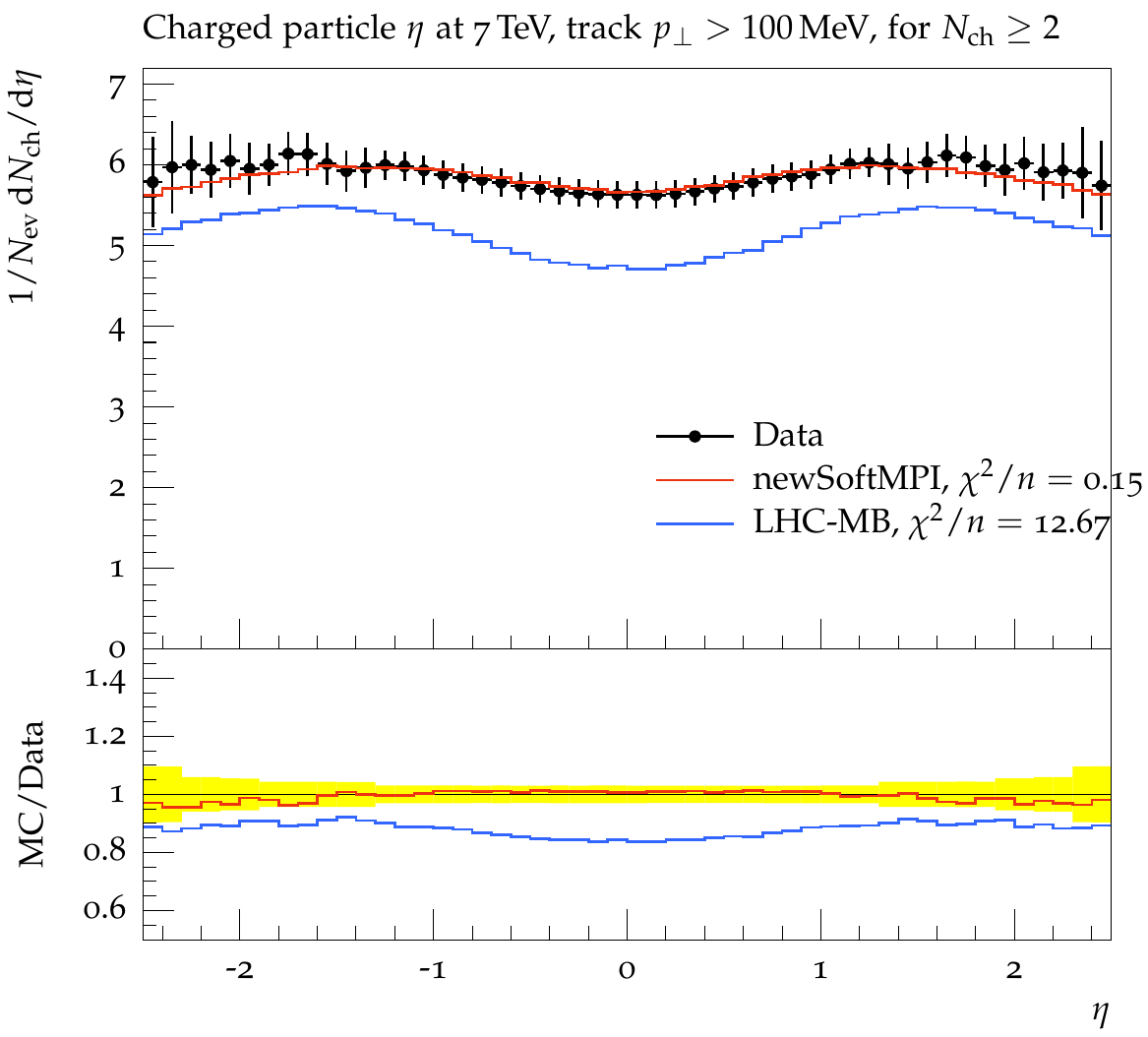}
  \caption{Pseudo-rapidity distribution in minimum-bias interactions as
    measured by ATLAS \cite{Aad:2010ac}, compared to simulations with the
    old and new MB models in Herwig. }
  \label{fig:etamb}
\end{figure}

\begin{figure}
  \centering
  \includegraphics[width=.49\textwidth]{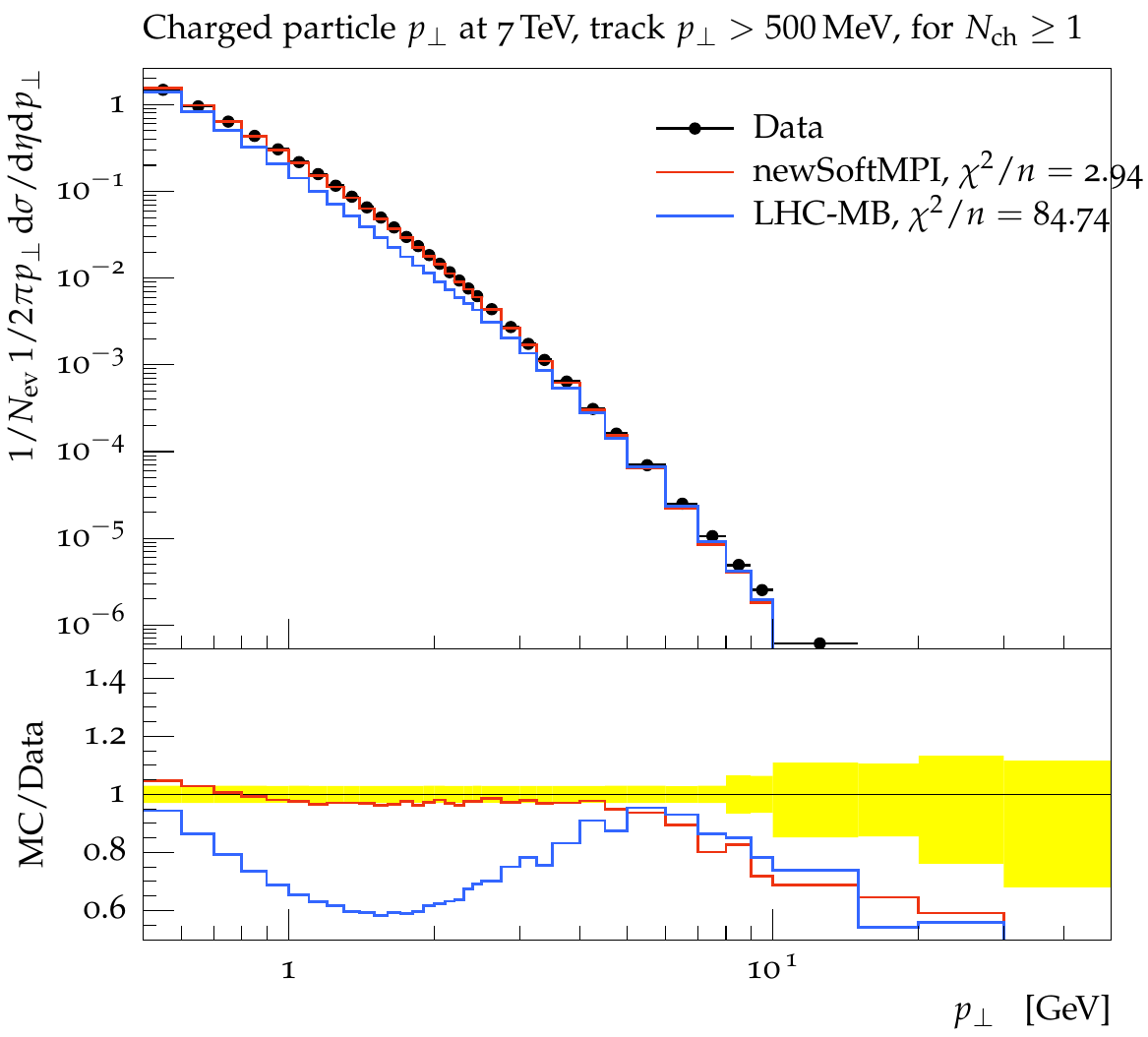}
  \hfill
  \includegraphics[width=.49\textwidth]{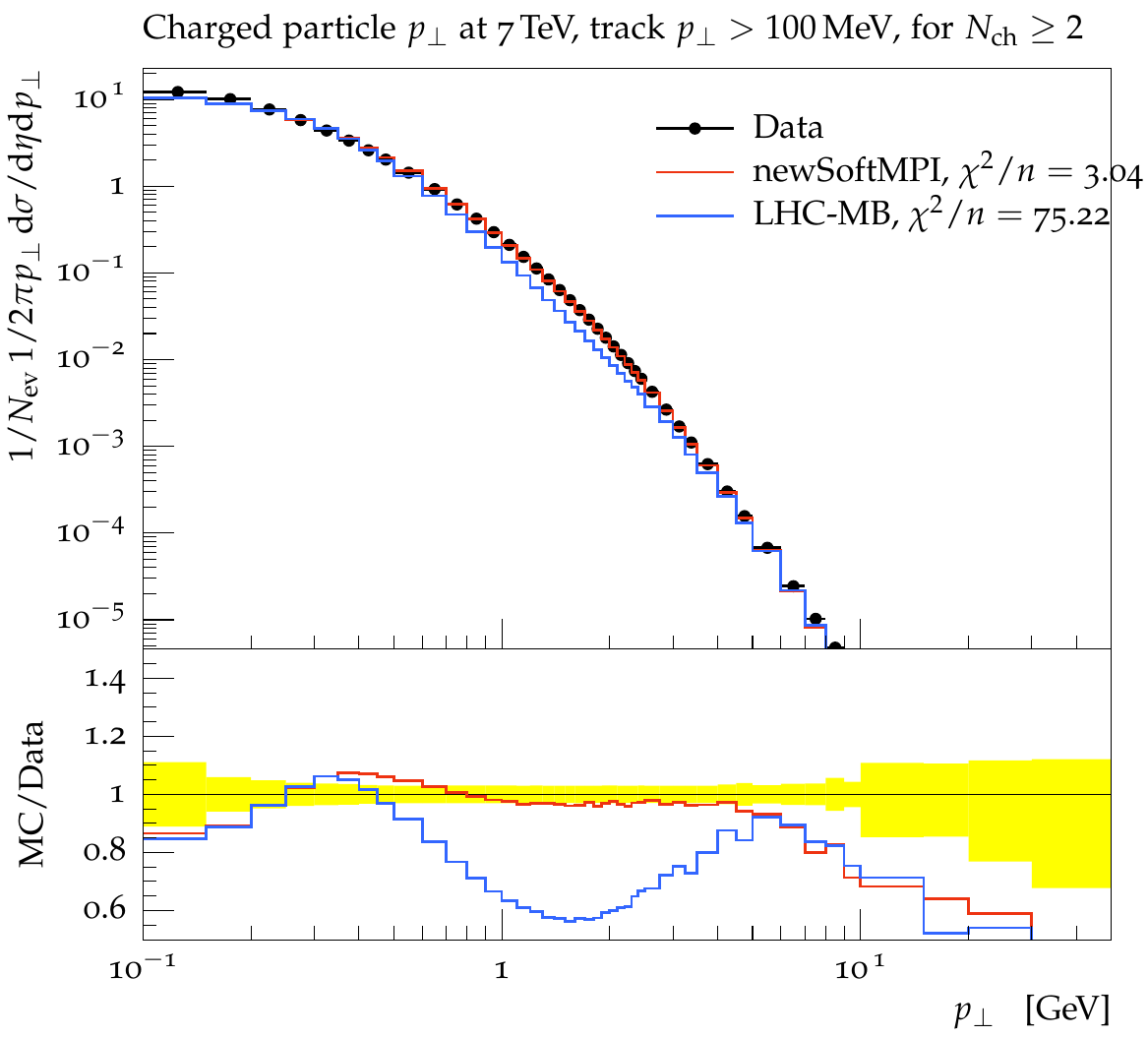}
  \caption{Transverse momentum distributions in MB interactions as
    measured by ATLAS \cite{Aad:2010ac}, compared to simulations with the
    old and new MB models in Herwig.}
  \label{fig:ptmb}
\end{figure}

Some results of these tunes are shown in Fig.~\ref{fig:etamb}.  We find
that the rapidity distribution of charged particles in minimum-bias
events is very well reproduced also for events with a significant
fraction of diffraction, the data is taken from ATLAS \cite{Aad:2010ac}.
These observables have been used to tune the model parameters.  In
Fig.~\ref{fig:ptmb} we show transverse momentum distributions of charged
particles in minimum-bias events that are also very well described in
the regime of soft particle production as measured by ATLAS
\cite{Aad:2010ac}.  Our tune has been focused on this regime and the
extension of this plot into the harder regime has to be properly
adjusted for in the semi-hard events from the MPI that have not been
adjusted in our study.  Once again, the results turn out very nicely and
show that we have much improved the description of this observable.

\begin{figure}
  \centering
  \includegraphics[width=.8\textwidth]{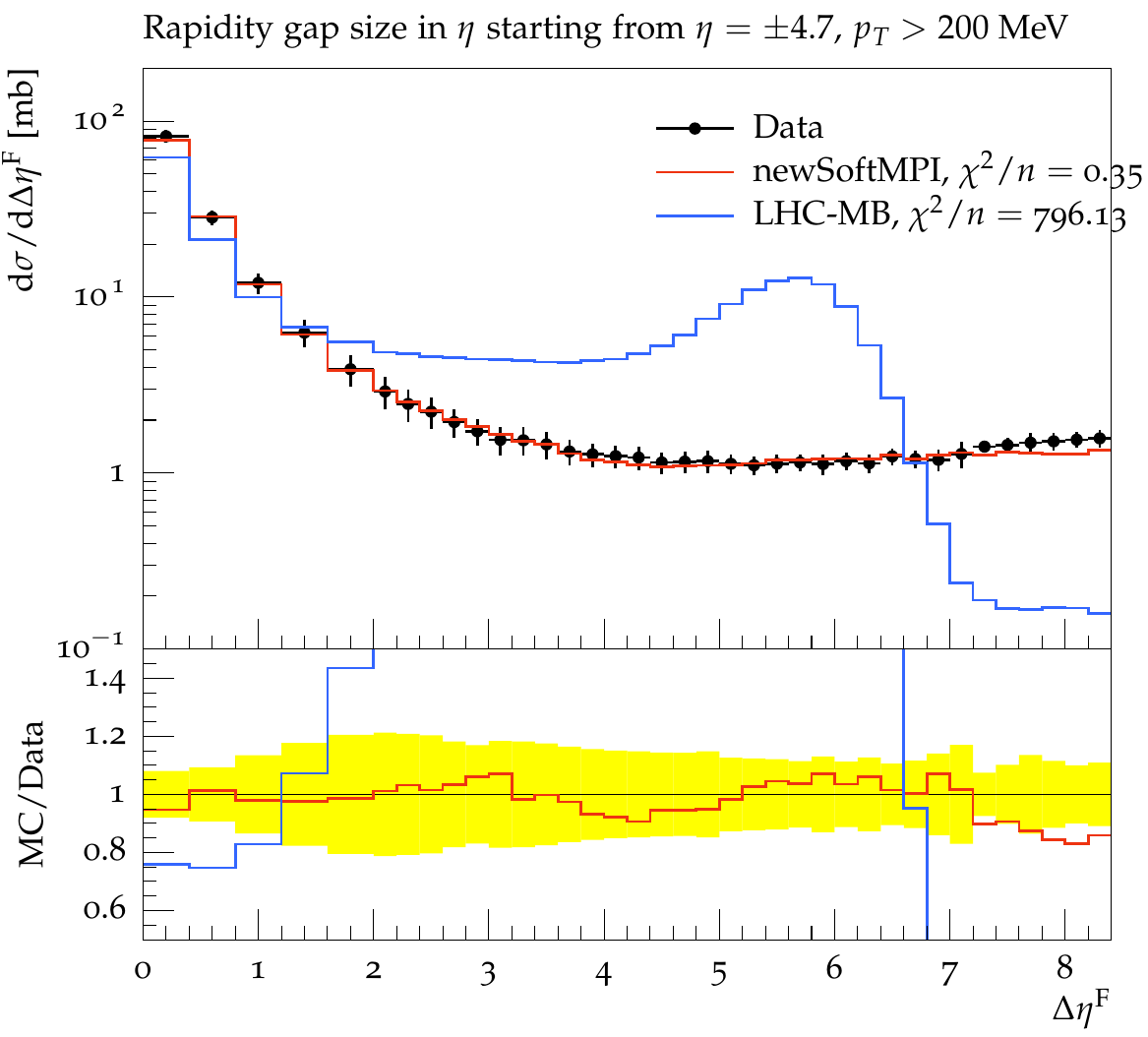}
  \caption{The $\Delta\eta^F$ distribution as measured by CMS
    \cite{Khachatryan:2015gka}, compared to the new model of soft
    interactions in Herwig.  }
  \label{fig:bumpnew}
\end{figure}

\begin{figure}
  \centering
  \includegraphics[width=.8\textwidth]{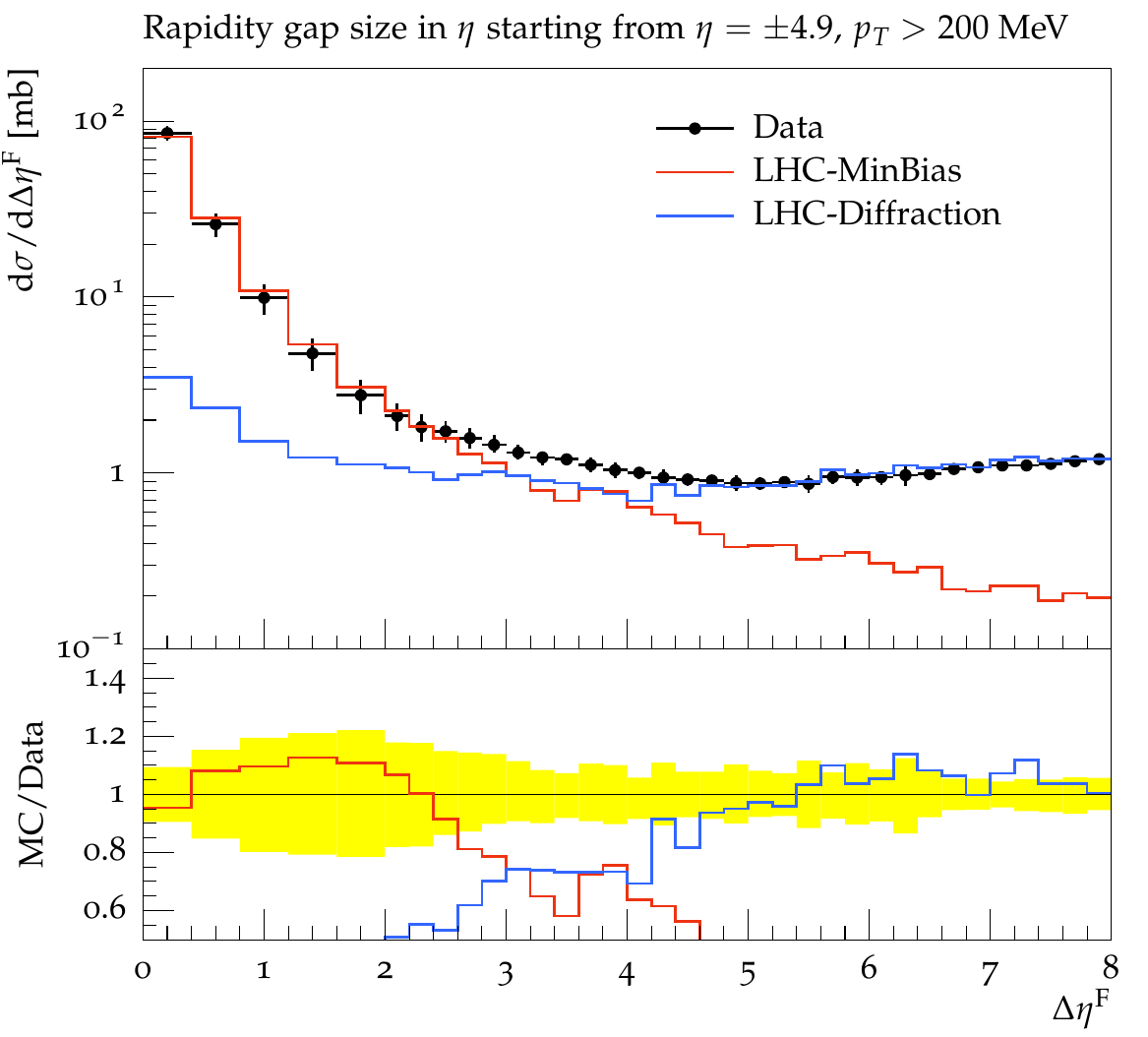}
  \caption{The diffractive and non-diffractive contributions to the
    $\Delta\eta^F$ distribution compared to ATLAS data
    \cite{Aad:2012pw}.  }
\label{fig:bumpcontribs}
\end{figure}

Last but not least, we consider the rapidity gap fraction measurement
that has initially motivated the extension of our model.  With this
observable we have adjusted the relative weight of the diffractive vs.\
non-diffractive contributions to the total cross section.  In
Fig.~\ref{fig:bumpnew} we show the result in comparison to CMS data
\cite{Khachatryan:2015gka}.  We find that with the new model the bump
disappears, as we do not have any more artificial rapidity gaps as a
result of unphysical colour assignments to the soft final states.  In
Fig.~\ref{fig:bumpcontribs} we also show the individual contributions
from the diffractive matrix elements vs the hard and soft MPI
contribution, called MinBias in the plot.  The MPI contribution shows
the exponential-like fall off for small rapidity gaps which is hardened
a bit by the MPI interactions.  The diffractive contribution is fairly
flat throughout the considered range in $\Delta\eta_F$.  

\section{Conclusion}
We have presented the new model for soft interactions in Herwig.  The
model adds diffractive final states to the simulation of minimum-bias
events and allows for the production of rapidity ordered gluons with
very small transverse momentum.  We can improve all considered
observables related to minimum-bias and diffractive events significantly
with this model.  This model is only the first of a number of steps to
improve the modeling of soft physics in Herwig.  

This work was supported in part by the European Union as part of the FP7
Marie Curie Initial Training Network MCnetITN (PITN-GA-2012-315877)

\bibliographystyle{ws-rv-van}


\end{document}